\documentstyle[12pt]{article}
\newcommand{\ho}{{\HH_{(0)}}}
\newcommand{\hto}{{{\tilde \HH}_{(0)}}}
\newcommand{\DT}{{\tilde\D}}
\def\ftoday{{\sl  \number\day \space\ifcase\month 
\or Janvier\or F\'evrier\or Mars\or avril\or Mai
\or Juin\or Juillet\or Ao\^ut\or Septembre\or Octobre
\or Novembre \or D\'ecembre\fi
\space  \number\year}}    
\newcommand{\journal}[4]{{\em #1~}#2\,(19#3)\,#4;}

\newcommand{\hpa}{\journal {Helv. Phys. Acta}}

\newcommand{\ijmp}{\journal {Int. J. Mod. Phys.}}

\newcommand{\pr}{\journal {Phys. Rev.}}

\newcommand{\jmp}{\journal {J. Math. Phys.}}

\newcommand{\cmp}{\journal {Commun. Math. Phys.}}

\newcommand{\cqg}{\journal {Class. Quantum Grav.}}

\newcommand{\np}{\journal {Nucl. Phys.}}
\newcommand{\pl}{\journal {Phys. Lett.}}

\newcommand{\prep}{\journal {Phys. Rep.}}

\makeatletter
\@addtoreset{equation}{section}
\makeatother
\renewcommand{\theequation}{\thesection.\arabic{equation}}
\newcommand{\es}{\\[3mm]}

           \newcommand{\G}{\Gamma}
\renewcommand{\d}{\delta}         \newcommand{\D}{\Delta}
\newcommand{\e}{\varepsilon}

\newcommand{\m}{\mu}
\newcommand{\n}{\nu}
         
\newcommand{\p}{\psi}              
\newcommand{\r}{\rho}
\newcommand{\s}{\sigma}           
         
\newcommand{\f}{{\phi}}           
\newcommand{\vf}{{\varphi}}
\newcommand{\BB}{{\cal B}}

\newcommand{\HH}{{\cal H}}

\newcommand{\NN}{{\cal N}}
\newcommand{\OO}{{\cal O}}

\newcommand{\SS}{{\cal S}}

\newcommand{\na}{\nabla}
\newcommand{\xint}{\dint d^4x\;}
\newcommand{\sla}{\raise.15ex\hbox{$/$}\kern -.57em} 
\newcommand{\Sla}{\raise.15ex\hbox{$/$}\kern -.70em}

\def\LP{\displaystyle{\Biggl(}}
\def\RP{\displaystyle{\Biggr)}}
\newcommand{\lp}{\left(}\newcommand{\rp}{\right)}
\newcommand{\lc}{\left[}\newcommand{\rc}{\right]}
\newcommand{\lac}{\left\{}\newcommand{\rac}{\right\}}

\newcommand{\complex}{{\kern .1em {\raise .47ex
\hbox {$\scriptscriptstyle |$}}
    \kern -.4em {\rm C}}}
\newcommand{\real}{{{\rm I} \kern -.19em {\rm R}}}
\newcommand{\rational}{{\kern .1em {\raise .47ex
\hbox{$\scripscriptstyle |$}}
    \kern -.35em {\rm Q}}}
\renewcommand{\natural}{{\vrule height 1.6ex width
.05em depth 0ex \kern -.35em {\rm N}}}

\newcommand{\tr}{{\rm {Tr} \,}}
\newcommand{\half}{\dfrac{1}{2}}
\newcommand{\pa}{\partial}

\newcommand{\dpad}[2]{{\displaystyle{\frac{\partial #1}{\partial #2}}}}
\newcommand{\dfud}[2]{{\displaystyle{\frac{\delta #1}{\delta #2}}}}
\newcommand{\dfrac}[2]{{\displaystyle{\frac{#1}{#2}}}}
\newcommand{\dsum}[2]{\displaystyle{\sum_{#1}^{#2}}}   
\newcommand{\dint}{\displaystyle{\int}}

\newcommand{\twiddle}{\lower.9ex\rlap{$\kern -.1em\scriptstyle\sim$}}

\newcommand{\equ}[1]{(\ref{#1})}
\newcommand{\eq}{\begin{equation}}
\newcommand{\eqn}[1]{\label{#1}\end{equation}}
\newcommand{\eea}{\end{eqnarray}}
\newcommand{\eqa}{\begin{eqnarray}}
\newcommand{\eqan}[1]{\label{#1}\end{eqnarray}}
\newcommand{\ba}{\begin{array}}
\newcommand{\ea}{\end{array}}
\newcommand{\eqac}{\begin{equation}\begin{array}{rcl}}
\newcommand{\eqacn}[1]{\end{array}\label{#1}\end{equation}}

\begin{document}
%********************************************************
%%\setlength{\baselineskip}{6ex}
%\setlength{\textwidth}{18cm}
%\setlength{\textheight}{26cm}
%%\setlength{\oddsidemargin}{-1cm}
%%\setlength{\evensidemargin}{-1cm}
%\setlength{\topmargin}{-2cm}
%********************************************************
%********************************************************
%{\large %POUR PREPRINT
%*********************************************************************
\newcommand{\cb}{{\bar c}}
\newcommand{\mn}{{\m\n}}
\newcommand{\pic}{$\spadesuit\spadesuit$}
\newcommand{\?}{{\bf ???}}

\titlepage  \noindent
{hep-th/9605158\\
UGVA-DPT-1996/05-925 \vspace{8mm}

\begin{center} 
{\huge{\bf Note on Constrained \\[2mm] 
              Cohomology}}
%              $\,^{^{\mbox{\large{\bf 1}}}}$ 
{\Large\footnote{Supported in part by the Swiss National 
Science Foundation and by OFES contract 93.0083 and Human Capital 
and Mobility, EC contract ERBCHRXCT920069}}
\vspace{8mm}

{\Large 
Fran{\c c}ois Delduc\footnote{Laboratoire de Physique Th\'eorique ENSLAPP,
URA 14-36 du CNRS, associee \`a l'ENS de Lyon, \`a l'universit\'e de Savoie et 
au LAPP, groupe de Lyon, ENS Lyon, 
All\'ee d'Italie 46,  F-69364 Lyon, France}
Nicola Maggiore\footnote{D\'epartement de Physique Th\'eorique, Universit\'e 
de Gen\`eve, quai E. Ansermet 24, CH-1211 Gen\`eve 4, 
Switzerland}$^,$\footnote{On leave of absence from Universit\`a degli 
Studi di Genova, Dipartimento di Fisica, Italy}
Olivier Piguet\footnote{Instituto da F{\'\i}sica, 
Universidade Cat\'olica de Petr\'opolis,
25610-130 Petr\'opolis, RJ, Brazil and 
Centro Brasileiro de Pesquisas F{\'\i}sicas (CBPF),
Rua Xavier Sigaud 150, 22290-180 Urca, RJ, 
Brazil}$^,$\footnote{On leave of absence from D\'epartement de 
Physique Th\'eorique, Universit\'e de Gen\`eve, Switzerland. Supported in 
part by the Brazilian National Research Council (CNPq)}\\
and Sylvain Wolf$\,$\footnotemark[3] %$^{\,3}$ 
}

\vspace{15mm}

\end{center}

\begin{center}
{\bf Abstract}
\end{center}
{\it
The cohomology of the BRS operator corresponding to a group of rigid
symmetries is studied in a space of local field functionals
subjected to a condition of gauge invariance. 
We propose a procedure
based on a filtration operator counting the degree in the
infinitesimal parameters of the rigid symmetry transformations.
An application to Witten's topological Yang-Mills theory is given.
}

\vfill\noindent
{\bf PACS codes:} 11.15.-q (gauge field theories), 03.65.Fd (algebraic methods),
03.70 (theory of quantized fields) 
\newpage
%*****************************************************
\section{Introduction}
All the symmetries of a given quantum field theory, including 
possibly the gauge invariance, may be grouped
together in a single extended BRS operator $D$~\cite{dixon}. 
 Here, whereas the ghosts associated to the gauge symmetry are 
represented by local fields, those associated with rigid 
symmetries are constant\footnote{See~\cite{bbbcd} 
for an earlier example of such a treatment for a rigid 
symmetry.}. Both types of ghosts have a Grassmann parity 
opposite to that of the corresponding generator. The proof of 
renormalizability is then a matter of computing the cohomology of $D$ in
the space of local field polynomials with dimensions restricted by power
counting~\cite{p-sor-book}. 

A typical example is provided by the supersymmetric Yang-Mills
theories quantized in the Wess-Zumino gauge. There, such a procedure
appears to be ne\-cessary~\cite{susy,mpw}
since  the supersymmetry generators 
do not form a closed algebra because of the presence of 
gauge transformations~\cite{bm}.

 Very often, it may be desirable to single out the individual
symmetries  from this unifying picture. This is needed, for instance, in
the construction of gauge invariant operators
belonging to some representation of a rigid symmetry.
This is a problem of constrained cohomology, i.e. of the
cohomology of $D$ in a space constrained by the requirement of
gauge invariance.

In the case of the super Yang-Mills theories,
the rigid symmetry is  the sup\-er\-sym\-metry itself, and an
important set of gauge invariant operators is 
 given by the components of the 
supercurrent multiplet~\cite{f-z-sc}, i.e. the supermultiplet of
currents which includes the $R$-axial current, the spinor current and
the energy-momentum tensor. 
The constrained cohomology approach we want to describe here 
was introduced~\cite{mpw} 
in this context, in order 
to construct the supercurrent~\cite{lmpw}.

It turns out that the structure involved is of the type studied
by Witten~\cite{wit} in the
framework of topological Yang-Mills theories. 

The method and its relevance for quantum
field theory are described in Section~\ref{covariance}. 
In particular, we reproduce the algebraic structure, called ``BRS 
extension'', found by Henneaux~\cite{henneaux} in a Hamiltonian
approach.

In Section~\ref{application} we briefly discuss the example of 
Witten's topological Yang-Mills theory. The reader may look 
at~\cite{mpw} for the supersymmetric gauge theoretical example.

%*****************************************************
\section{Covariance of Gauge Invariant Operators Under 
  a Group of Rigid Symmetries}\label{covariance} 
The quantization of gauge theories possessing a rigid symmetry
is conveniently performed by introducing an extended 
nilpotent BRS operator $D$ which, besides the BRS operator $s$
associated with gauge invariance, 
also includes the infinitesimal transformations of the rigid 
symmetry~\cite{susy,mpw}.
Let us write the infinitesimal transformations of the fields $\vf^i$, in
the classical theory, as\footnote{Summation over repeated indices is
always understood.} 
\eq
\d_{\rm rigid}\,\vf^i :=\e^A\d_A\vf^i\ ,
\eqn{rigid-trf}
with, in the most general case\footnote{In the case of a
superalgebra, the bracket $[\cdot,\cdot]$ is a graded commutator, i.e.
an anticommutator if both its arguments are fermionic, 
and a commutator otherwise.
We shall keep this notation throughout the paper.} 
\eq
\lc \d_A,\d_B \rc = -c_{AB}{}^C \d_C
  \quad + \mbox{ eqs. of motion + gauge transf.}  \ ,
\eqn{alg-delta}
\eq
\lc \d_A, s \rc = 0 \ .
\eqn{alg1-delta}
$\d_{\rm rigid}$ may act nonlinearly on the fields, and the algebra may
happen to
close only modulo equations of motion and 
field-dependent gauge transformations
\footnote{In the 
special case where the action of $\d_A$ 
is linear:
\[
\d_A\vf^i = T_A{}^i{}_j\vf^j\ ,
\]
and the algebra closes off-shell, this corresponds to the matrix algebra
\[
\lc T_A,T_B \rc = c_{AB}{}^C T_C\ .
\]}.
The extended BRS operator is  then defined by 
\eq
D:= s + \e^A\d_A +\half c_{AB}{}^C \e^A\e^B\dpad{}{\e^C}  + O(\e^2)\ ,
\eqn{def-D}
where the infinitesimal parameter $\e^A$ is a constant ghost: it is 
an anticommuting (resp. commuting) number if 
the generator $\d_A$ is a bosonic (resp. fermionic) operator. 
The term involving the structure constants $c_{AB}{}^C$ ensures the
nilpotency of $D$:
\eq
D^2 = 0  \quad\mbox{(modulo eqs. of motion)} \ ,
\eqn{D-nilpotency}
 which expresses in a compact way the whole 
algebra of infinitesimal generators.
Terms quadratic in $\e$ are present in \equ{def-D}
if the rigid algebra \equ{alg-delta}
closes modulo field dependent gauge 
transformations, as  it is the case, 
for instance, in the supersymmetric gauge theories in the Wess-Zumino 
gauge~\cite{susy,mpw}.

The invariance of the quantum theory under the extended BRS transformations
\equ{def-D},  i.e. under the gauge BRS symmetry $s$ and the rigid symmetry
$\d_A$, is expressed by a Slavnov-Taylor identity -- we assume that
there is no anomaly --
\eq
\SS(\G) := \dint dx\; \dsum{i}{}
    \dfud{\G}{\vf^*_i(x)}\dfud{\G}{\vf^{i}(x)}
    + \half c_{AB}{}^C\e^A\e^B\dpad{\G}{\e^C} = 0\ .
\eqn{slavnov}
Here $\G(\vf,\vf^*)$ is the vertex functional, i.e. the generating
functional of the 1-particle-irreducible, amputated Green functions, and
the $\vf^*_i$ are external fields coupled to the extended BRS
transformations of the fields $\vf^i$, introduced at the classical 
level by adding to the invariant action the terms
\eq
S_{\rm ext} = \dint dx\; \dsum{i}{} \vf^*_i D\vf^i +O((\vf^*)^2)\ .
\eqn{S-ext}
The terms quadratic in the external sources are necessary if the whole
algebra, as expressed by 
the nilpotency of $D$, holds only on-shell~\cite{bv}.

The Slavnov-Taylor identity also defines the quantum form of the field
transformations. The extended BRS transformations of a
local quantum composite field operator $\OO$, 
whose vertex functions are generated by the insertion functional
$\OO\cdot\G(\vf,\vf^*)$, is defined by 
\eq
\BB_\G (\OO\cdot\G) \ ,
\eqn{bb-insertion}
where the $\G$-dependent linear functional operator $\BB_\G$ is 
the linearized Slavnov-Taylor operator
\eq
\BB_\G := \dint dx\; \dsum{i}{}
    \LP \dfud{\G}{\vf^*_i(x)}\dfud{}{\vf^{i}(x)}
    +\dfud{\G}{\vf(x)}\dfud{}{\vf^*_i(x)}\RP
    + \half c_{AB}{}^C\e^A\e^B\dpad{}{\e^C} \ .
\eqn{slavnov-lin}
This operator is automatically nilpotent:
\eq
(\BB_\G)^2=0\ ,
\eqn{nilpot}
if $\G$ obeys the Slavnov identity \equ{slavnov}.
The action of $\BB_\G$ on the fields $\vf$
and $\vf^*$ defines their extended BRS transformations at the quantum level.

The problem we want now to solve is: how to extract the
gauge-BRS and the rigid transformation 
laws of the quantum fields, together with their algebra,
from the nilpotent extended BRS transformations 
containing all the symmetries of the theory ? 

The answer begins by introducing a filtration operator 
\eq
\NN := \e^A\dpad{}{\e^A}\ ,
\eqn{filtration}
whose eigenvalues are the nonnegative integers, 
its application being restricted
to the polynomials in $\vf$, $\vf^*$ and $\e$.
The vertex functional and the quantum BRS operator \equ{slavnov-lin}
 may be expanded
according to these eigenvalues:
\eq
\G = \dsum{n\ge0}{} \G_n\ ,\quad
\BB_\G = \dsum{n\ge0}{} \BB_n\ ,
\eqn{exp-gamma-bb}
where
\eq
\NN\G_n = n\G_n\ ,\quad \lc \NN,\BB_n \rc = n\BB_n\ .
\eqn{eigenval-eq}
The nilpotency of $\BB_\G$  implies, at order 0, 1 and 2, the 
equations
\eq\ba{l}
(\BB_0)^2 = 0\ ,\es
\lc\BB_0,\BB_1\rc = 0\ ,\es
(\BB_1)^2 + \lc \BB_0,\BB_2 \rc = 0\ . 
\ea\eqn{bb-n-alg}
$\BB_0$ is interpreted as the 
gauge-BRS operator. It is natural to interpret $\BB_1$ 
as the operator of rigid transformations. More precisely,
we can write
\eq
\BB_1 =: \e^A X_A\ ,\quad \BB_2 =: \half\e^A\e^B X_{AB}\ .
\eqn{def-generators}
The first equation  defines the generators of the 
rigid transformations we were looking for. 
The second equation defines ``second order'' transformations.
The algebra of the gauge-BRS and rigid operators is easily deduced from
\equ{bb-n-alg}: 
\eq\ba{l}
(\BB_0)^2 =0 ,\es
\lc \BB_0,X_A \rc = 0\ ,\es
\lc X_A,X_B \rc + c_{AB}{}^C X_C + \lc \BB_0,X_{AB} \rc = 0\ .
\ea\eqn{alg-generators}
One sees that the rigid generators $X_A$ do not fulfill
in general a closed (super)Lie algebra\footnote{ The third of eqs.
\equ{alg-generators} precisely represents the algebraic structure one
encounters in supersymmetry~\cite{mpw,bm}, as mentioned
in the introduction.}. 
But this will however be the case
if we restrict their action to the space of gauge invariant operators. 
A gauge invariant local field operator $[\OO]$
is defined as a cohomology class of invariant composite insertion
$\OO\cdot\G$ (c.f. \equ{bb-insertion}) :
\eq%\ba{l}
%\mbox{(1)}\quad
\BB_0(\OO\cdot\G) = 0 ,%\es
%\mbox{(2)}\quad\mbox{There is no insertion }\hat\OO\cdot\G \mbox{ such that }
%   \OO\cdot\G = \BB_0 \lp \hat\OO\cdot\G\rp\ .
%\ea
\eqn{g-inv-op}
with 
\eq
[\OO]=0 \ \ \ \mbox{if}\ \ \ \OO\cdot\G = \BB_\G (\hat\OO\cdot\G) \ \ 
\forall\, \OO \in [\OO]\ .
\eqn{class}
%This defines a gauge invariant operator as a 
%$\BB_0$-cohomology class, denoted by  $[\OO]$. 
There is a natural definition of the action of the rigid generators
$X_A$ on these cohomology  classes, since these generators
commute with the
gauge-BRS operator (see \equ{alg-generators}). It is then clear that
\eq
(\BB_1)^2 [\OO]= [0]\ ,
\eqn{equiv-nilpot}
or, equivalently,
\eq
\lp \lc X_A,X_B \rc + c_{AB}{}^C X_C \rp [\OO]=[0]\ ,
\eqn{alg-on-g-inv}
on any gauge invariant operator $[\OO]$. 
%In \equ{equiv-nilpot} and in \equ{alg-on-g-inv}, 
%$[0]$ denotes the cohomology class of the null element.

\noindent{\bf Remark:\ } The nilpotency 
condition of $\BB_1$ on the gauge invariant operators
may be called a constrained  nilpotency condition. The computation of the
cohomology of $\BB_1$ in the space of the gauge invariant operators
is known as a problem of invariant cohomology~\cite{stora}, or
``constrained cohomology''.
We shall apply the above formalism in the next section
to a simple derivation of Witten's
observables~\cite{wit}
in four-dimensional topological Yang-Mills theory.

%*****************************************************
\section{Application: Topological Yang-Mills Theory}\label{application}
Witten's topological
Yang-Mills field theory~\cite{wit}
in $D=4$  Euclidean space is described by 
the following set of fields: a gauge field $A_\m$ with its associated
ghost $c$, and a fermionic vector field $\p_\m$ with its associated ghost
$\vf$. %The ghost numbers and Grassmann parity of these fields are
%shown in Table~\ref{table1}.
%****************************************************

% TABLE A LA FIN
%**********************************************************
These fields take their value in the adjoint representation of the 
gauge group, chosen as an
arbitrary compact Lie group with structure 
constants\footnote{We use a matrix notation, e.g. $A_\m=A_\m^a\tau_a$,
with $[\tau_a,\tau_b]=f_{ab}{}^c\tau_c$, and with the trace 
normalized in such a way that
$\tr \tau_a\tau_b=\d_{ab}$.} $f_{ab}{}^c$.

The invariances of this theory, namely 
the gauge symmetry and a sup\-er\-sym\-met\-ric-like shift symmetry,
are grouped together in an extended BRS operator $D$, whose action on
the fields is given by
\eq\ba{l}
DA_\m = -\nabla_\m c + \e\p_\m\ ,\es
D\p_\m = [c,\p_\m] -\e \nabla_\m\vf\ ,\es
D\vf = [c,\vf]\ ,\es
Dc = c^2 -\e^2\vf\ ,\es
D\e= 0\ ,
\ea\eqn{t-ext-brs}
with
\eq
D^2=0\ .
\eqn{t-nilpotency}
$\nabla_\m$ is the covariant derivative: $\nabla_\m\cdot =
\pa_\m\cdot+[A_\m,\cdot]$. 
The supersymmetry infinitesimal parameter
$\e$ is taken to be commuting and plays the role of a constant 
ghost. Usually, $\e$ is absorbed in a redefinition of the fields
$\p_\m$ and $\vf$, which accordingly acquire ghost numbers 1 
and 2, respectively. We shall nevertheless keep $\e$ explicit 
to better illustrate our procedure. We remark that the presence of the
term quadratic in $\e$ in the transformation law of the ghost field $c$
is necessary for the nilpotency of $D$.
In this parametrization all the fields 
introduced up to now have ghost number 0 except $c$
and $\e$, of ghost number 1. 

 The classical action reads~\cite{wit}
\eq
S =  \dfrac{1}{4g^2}\tr\xint \lp  F^{+\mn}F^+_{\mn} 
  +\cdots  \rp\ ,
\eqn{action}
where
\eq
F^+_{\m\n}=\half\lp F_{\m\n}-\half\e_{\m\n\r\s}F^{\r\s} \rp \ ,
\eqn{antiselfdual}
is the anti-selfdual part of the Yang-Mills field 
strength. The dots represent the terms needed for the gauge 
fixing~\cite{osb,bmops}.

This theory is renormalizable, and even ultraviolet finite. 
The proof~\cite{bmops} is based on the triviality of the 
cohomology of the extended BRS operator $D$. However,
in order to maintain the
exposition simple, we shall keep ourselves at the level
of the classical theory
described by the action \equ{action}, the generalization to the quantum
theory being straightforward.

The filtration operator \equ{filtration} 
and the expansion \equ{exp-gamma-bb} take here the form
\eq
\NN = \e\dpad{}{\e}\ ,
\eqn{t-filtration}
\eq
D = D_0+ \e D_1 + \e^2 D_2\ ,
\eqn{t-expansion}
where we factorized the powers of $\e$. 
The transformation laws of the operators $D_n$ and their algebra 
-- up to order 2 -- read
\eq\ba{lll}
D_0 A_\m = -\nabla_\m c\quad  &D_1 A_\m= \p_\m\ ,\quad 
&D_2A_\m=0 \ ,\es
D_0\p_\m = [c,\p_\m]\ ,\quad  &D_1\p_\m = - \nabla_\m\vf\ ,\quad
&D_2\p_\m=0\ ,\es
D_0\vf = [c,\vf]\quad         &D_1\vf = 0\ ,\quad 
&D_2\vf = 0\ ,\es
D_0 c = c^2\ ,\quad           &D_1 c = 0\ ,
&D_2 c = -\vf \ ,\es
\ea\eqn{t-filt-brs}
\eq
(D_0)^2 = 0\ ,\quad [D_0,D_1]= 0\ ,
\quad (D_1)^2 + [D_0,D_2] = 0\ .
\eqn{t-algebra}

Following the lines drawn in the previous section for the general case, 
we consider the space $\ho$
of gauge invariant local operators, i.e. the cohomology of $D_0$. 
Within this space, the analogous of~\equ{equiv-nilpot} reads
\eq
D_1^2 \D = 0\ ,\quad\forall\, \D\in\ho 
\eqn{equiv-d1}
which must be intended according to cohomology classes. 

As it is shown in the Appendix, it follows from
the transformation laws \equ{t-filt-brs} that the 
the constrained cohomology of $D_1$, i.e. the cohomology of $D_1$ in
$\ho$,
%which we denote with ${\cal H}_{inv}(D_1)$,  
is given by the $D_0$-cohomology classes represented by the functions
\eq
F\lp P_{\rm inv}(c),\,Q_{\rm inv}(\vf)\rp\ ,
\eqn{cohom-d1}
whose arguments are group invariant polynomials
which depend only on the fields $c$ and $\vf$ without their derivatives.
%BRS invariant polynomials in the undifferentiated field $\vf$ and in the
%ghost $c$ with all its derivatives :
%\eq
%{\cal H}_{inv}(D_1) = P(\vf,c,\pa_\m c,\pa_\m\pa_\n c,\ldots)\ .
%\eqn{cohom-d1}
%The easiest way to see that, 
%following \cite{p-sor-book}, is to observe that the fields $A_\m$, $\psi_\m$ 
%and $\pa_\m\phi$ form a kind of triplet of the form $s a =b$, $s b=c$, $sc=0$, 
%and therefore they do not belong to the constrained cohomology of $D_1$, 
%nor do their derivatives.

The result \equ{cohom-d1} shows that, to 
the contrary of the cohomology of $D$, the constrained cohomology 
is not empty. The filtration we introduced in the global ghosts $\e$
allowed to see this easily. 

According to the definition of \cite{wit}, the observables $\OO$ 
of the model should be independent of $c$, thus excluding for instance
terms like $\tr c^{2n+1}$.
The consequence of the $c$-independence is that the elements of the 
constrained space in which calculating the cohomology of $D_1$
are invariant under $D_2$. This property, together with the third of 
eqs.~\equ{t-algebra}, implies the $exact$ nilpotency of $D_1$ (and not in 
the sense of the cohomology classes):   
\eq
D_0\OO=D_2 \OO =0 \rightarrow (D_1)^2 \OO =0 \ .
\eqn{properties}                                           
This means that 
the zero ghost cohomology of $D$ in $\ho$ is identical to that of $D_1$. 
From \equ{cohom-d1}, requiring 
$c$-independence, we recover the Witten's observables of the 
topological Yang-Mills theory: 
\eq
\lac\OO_{\rm Witten}\rac = \lac Q_{\rm inv}(\vf) \rac\ ,
\eqn{wit-obs}
which are invariant polynomials of the field $\vf$ only, 
with the exclusion of its derivatives.
As a comment, we remark that the above result \equ{wit-obs} 
has been obtained after a simple cohomology
calculation in a restricted functional space (see the Appendix), 
while usually the Witten's 
observables are characterized as what is called the ``equivariant'', 
or ``basic'', cohomology~\cite{equiv}, which requires a specific,
and more complicated, analysis~\cite{equiv1}.

%*******************************************************
\section{Conclusions}
We have been able to extract the individual symmetries, namely the gauge
invariance and the rigid symmetries, from the general BRS operator of
the theory. We have shown in particular that, although the generators
constructed in this way do not form a closed algebra, they do so if
their action is restricted to the space of gauge invariant quantities
defined by the cohomology of the gauge BRS operator. 
Only in this 
restricted space it makes sense to study the
cohomology corresponding to the rigid invariance, and we stress
that this holds for any
gauge theory characterized by an additional rigid invariance satisfying
the general algebra \equ{alg-delta}.
As an example, we
recovered the Witten's observables of the topological Yang-Mills theory, 
which are usually found in the more complicated context of the 
equivariant cohomology.
Moreover, the formalism proposed in this paper is well adapted
to the quantum theory, since the constraints such as gauge invariance,
as well as the symmetry generators,
are expressed in a functional way, i.e. as constraints and operations 
on the generating functionals of Green functions.

%**************************************************************
\vspace{12mm}

\noindent   {\bf Acknowledgments}: One of the authors 
(O.P.) is very indebted to the Conselho Nacional de Pesquisa e Desevolvimento  
(Brazil)  for its financial support,  to the Institute of Physics of 
the Catholic University of Petr\'opolis (Brazil) and its head Prof. R. Doria,
as well as to the CBPF (Rio de Janeiro),
where part of this work has been done, for their hospitality.
We are grateful for interesting discussions with 
M.A. de Andrade, C. Becchi, O.M. del Cima, N. Dragon and S.P. Sorella on
this work, and
we are glad to thank R. Stora and M. Henneaux 
for a critical reading of the 
manuscript and for drawing our attention to refs.~\cite{equiv} and
\cite{henneaux}.

%*****************************************************************
%****************************************************************
%\appendix
\renewcommand{\theequation}{\Alph{section}.\arabic{equation}}
\renewcommand{\thesection}{\Alph{section}}
\setcounter{equation}{0}
\setcounter{section}{1}
\section*{Appendix}
As said in Sect. \ref{application}, the operator $D_1$ is nilpotent
within the space $\ho$ of the local
cohomology classes of the gauge BRS operator
$D_0$ (see \equ{equiv-d1}). This space is isomorphic~\cite{local-coho}
to the space $\hto$
of the gauge invariant polynomials generated by the invariant
polynomials $P_{\rm inv}(c)$ and $Q_{\rm inv}(F,\p,\vf)$, the former
depending on the field $c$ but not on its derivatives, 
the latter depending on the fields $F_{\m\n}$, $\p_\m$, $\vf$ and all
their covariant derivatives. 
One notes that
\eq
(D_1)^2\DT = 0\ ,\quad\forall\, \DT\in\hto\ ,
\eqn{nilp-d1}
%due to the invariance of $P_{\rm inv}(c)$. This, together with the
and 
\eq
D_1\hto \subset \hto\ .
\eqn{inclusion}
Thus $D_1$ is a coboundary operator in $\hto$.

The isomorphism of $\ho$ and $\hto$ implies that 
the image of $D_1$ in $\ho$ is isomorphic to its image in $\hto$: 
\eq
{\rm Im}\,D_1\vert_\ho \approx {\rm Im}\,D_1\vert_\hto\ .
\eqn{images}
In order to see this, let us consider an arbitrary representative 
$\DT+D_0(\cdots)$ of an element $\D$ of
$\ho$ represented by $\DT\in\hto$. Applying $D_1$ to it we get
\[ 
D_1\lp\DT+D_0(\cdots)\rp = D_1\DT - D_0D_1(\cdots)\ ,
\]
due to the anticommutativity of $D_0$ and $D_1$, i.e. its image is
represented by $D_1\DT$, which belongs to $\hto$ due to the 
property \equ{inclusion}.

Moreover, it also follows from the latter property that 
the kernel of $D_1$ in $\ho$ is 
isomorphic to its kernel in $\hto$:
\eq
{\rm Ker}\,D_1\vert_\ho \approx {\rm Ker}\,D_1\vert_\hto\ .
\eqn{kernels}
Indeed, the $D_1$-invariance of an element of $\ho$ represented by  
$\DT\in\hto$ is expressed on any of its representatives by the condition
\[
D_1\lp \DT + D_0(\cdots) \rp = D_0(\cdots)\ ,
\]
which means that $D_1\DT=D_0(\cdots)$, i.e. $D_1\DT=0$ since $\hto$ does
not contain $D_0$-exact elements.

The results \equ{images} and \equ{kernels} show that the
cohomologies of $D_1$ in $\ho$  and in $\hto$ are isomorphic. We are
thus left to compute the latter cohomology.

From the observation that
\[\ba{l}
D_1\lp\na_{\m_1}\cdots\na_{\m_n}F_{\m\n}\rp = \\
\quad   \na_{\m_1}\cdots\na_{\m_n} \lp\na_\m\p_\n-\na_\n\p_\m\rp
  +\dsum{i=1}{n} \na_{\m_1}\cdots\na_{\m_{i-1}}
       \lc\p_{\m_i}\,,\,\na_{\m_{i+1}}\cdots\na_{\m_n}F_{\m\n}\rc\ ,\\[5mm]
D_1\lp\na_{\m_1}\cdots\na_{\m_n} \lp\na_\m\p_\n+\na_\n\p_\m\rp\rp =\es
\quad   \na_{\m_1}\cdots\na_{\m_n}
    \lp-\na_\m\na_\n\vf-\na_\n\na_\m\vf+2\lc\p_\m\,,\,\p_\n\rc\rp \\
\quad    +  \dsum{i=1}{n} \na_{\m_1}\cdots\na_{\m_{i-1}}
       \lc\p_{\m_i}\,,\,\na_{\m_{i+1}}\cdots\na_{\m_n}
          \lp\na_\m\p_\n+\na_\n\p_\m\rp\rc\ ,\\[5mm]
\lc\na_\m,\na_\n\rc\f = \lc F_{\m\n}\,,\,\f  \rc\ ,\quad\forall\,\f\ ,
\ea\]
we see that we can choose, as independent variables, the fields
\eq
\lac X,\ D_1X,\ \vf,\ c\rac\ ,
\eqn{ind-variables}
with
\[\ba{l}
\hspace{-10mm} X :=\\
F_{\m\n}\ \mbox{and its symmetric covariant derivatives}\ ,\\ 
\na_\m\p_\n+\na_\n\p_\m \ \mbox{and its symmetric 
           covariant derivatives}\ ,\\
\p_\m \ .
\ea\]
In this basis the BRS operators read
\[\ba{l}
D_0 = \dsum{X}{}\lp [c,X]\dpad{}{X} + [c,D_1X]\dpad{}{(D_1X)}
   + [c,\vf]\dpad{}{\vf} + c^2\dpad{}{c} \rp\ ,\es
D_1 = \dsum{X}{}\lp D_1X\dpad{}{X} + [\vf, X]\dpad{}{(D_1X)} \rp\ 
,\es
D_2 = -\vf\dpad{}{c}\ ,
\ea\]
and obey the algebra \equ{t-algebra}. In particular $D_1$ 
is not nilpotent, when applied to the variables \equ{ind-variables},
although it does when applied to the elements of $\hto$.
Extending a well known argument~\cite{p-sor-book} to the present case of an
operator which is not nilpotent, we define the operators
\[
D'_1 := X\dpad{}{(D_1X)}\ ,\quad F := X\dpad{}{X} + D_1X\dpad{}{(D_1X)}\ ,
\]
we check that
\[
[D_0,F] = 0\ ,\quad [D_1,F] = 0\ ,\quad [D_1,D'_1] = F\ ,
\]
and we note that
\[
D'_1\hto \subset \hto\ .
\]
For solving the cohomology equation
\[
D_1\DT = 0\ ,\quad \DT \in \hto\ ,
\]
we expand $\DT$ according to the eigenvalues of the operator $F$:
\[
\DT = \dsum{n\ge0}{} \DT^{(n)}\ ,\quad
     \mbox{with}\quad F\DT^{(n)}=n\DT^{(n)}\ .
\]
We have then
\[
D_0\DT^{(n)}=0\ ,\quad D_1\DT^{(n)}=0\ ,\quad\forall\, n\ge0\ ,
\]
and, defining
\[
\DT' := \dsum{n\ge1}{}\dfrac{1}{n}D'_1\DT^{(n)}\ ,
\]
which belongs to $\hto$, we get
\[
\DT = \DT^{(0)} + D_1\DT'\ .
\]
This shows that the cohomology of $D_1$ in $\hto$ is given by
the $X$- and $D_1X$-independent $\DT^{(0)}$, i.e. that it consists 
of the elements of $\hto$ of the form \equ{cohom-d1}. This is the result
announced in Sect. \ref{application}, due to the isomorphism of 
the cohomologies of $D_1$ in $\hto$ and in $\ho$.


\begin{thebibliography}{999}
\bibitem{dixon} J.A. Dixon, \cmp{140}{91}{169}
\bibitem{bbbcd}  C. Becchi, A. Blasi, G. Bonneau, R. Collina and
                     F. Delduc, \cmp{120}{88}{121}
\bibitem{p-sor-book} O. Piguet and S.P. Sorella, ``Algebraic
                 Renormalization'', {\em Lecture Notes in Physics, 
                 Vol.} m28 
                 ({\em Springer Verlag, Berlin,
                 Heidelberg}, 1995);  
\bibitem{susy} P.L. White, \cqg {9}{92}{413}\\ \cqg {9}{92}{1663}\\
             N. Maggiore, \ijmp{A10}{95}{3781}\\ \ijmp{A10}{95}{3937} \\
            N. Maggiore, O. Piguet and M. Ribordy, \hpa{68}{95}{264} 
\bibitem{mpw} N. Maggiore, O. Piguet and S. Wolf, \np{B458}{96}{403}
\bibitem{bm}     P. Breitenlohner and D. Maison,
                 ``Renormalization of supersymmetric Yang--Mills 
                 theories'', {\em 
                 in Cambridge 1985, Proceedings:
                 ``Supersymmetry and its applications''}, 
                 p.309;\\
                 and ``N=2 Supersymmetric Yang--Mills theories
                 in the Wess--Zumino gauge'', {\em in
                 ``Renormalization of quantum field 
                 theories with
                 nonlinear field transformations'', 
                 ed.: P. Breitenlohner, D. Maison and K. 
                   Sibold, 
                 Lecture Notes in Physics, Vol.} 303, 
                  p.64 ({\em Springer Verlag, Berlin,
                 Heidelberg}, 1988);
\bibitem{f-z-sc} S. Ferrara and B. Zumino, \np{B87}{75}{207}
\bibitem{lmpw}  work in progress;
                %C. Lucchesi, N. Maggiore, O. Piguet and S. Wolf 
\bibitem{wit} E. Witten, \cmp{117}{88}{353}\\ \ijmp{A6}{91}{2775}
   %``Introduction to the Cohomological Field
    %    Theories'', {\em Lectures at the Workshop on Topological Methods in
     %       Physics, ICTP, Trieste, Italy} (1990);
\bibitem{henneaux} M. Henneaux, \np{B308}{88}{619}
\bibitem{bv}     I.A. Batalin and G.A. Vilkovisky, 
    \pl {69B}{83}{309}\\
     \pr {D28}{81}{2567} \pl {102B}{81}{27}
%\bibitem{moore} S. Cordes, G. Moore, S. Ramgoolam, ``Lectures 
%                 on 
%                 2-d Yang-Mills theory, equivariant cohomology and 
%                 topological field theories'', {\em Les Houches 
%                 1994}, hep-th/9411210;
\bibitem{stora} R.Stora, private communication;
\bibitem{bbrt} D. Birmingham, M. Blau, M. Rakowski and G. Thompson,\\
               \prep{209}{91}{129}
\bibitem{osb} S. Ouvry, R. Stora and P. van Baal, \pl{B220}{89}{159}
\bibitem{bmops}    M.W. de Oliveira, \pl{B307}{93}{347}\\
           A. Brandhuber, O. Moritsch, M.W. de Oliveira, O. Piguet
              and M. Schweda, \np{B431}{94}{173}
\bibitem{equiv} J. Kalkman, \cmp{153}{93}{447}\\
                R.Stora, ``Equivariant cohomology and topological theories'',
                Lecture given at the international symposium on BRS
                symmetry, September 1995, Kyoto, Japan, Enslapp-A-571/95;\\
                R. Stora, F. Thuillier and J.-C. Wallet, ``Algebraic 
                structure of cohomological field theory models and 
                equivariant cohomology'', Lectures at the 1st Caribbean 
                Spring School of Mathematics and Theoretical Physics, 
                Saint Fran{\c c}ois, Guadeloupe, 1993;
\bibitem{equiv1} S. Ouvry, R. Stora and P. van Baal, \pl{B220}{89}{159}\\
                 A. Blasi and R. Collina, {\it Phys. Lett.} B222(1989)419
\bibitem{local-coho} G. Bandelloni, \jmp{27}{86}{2551}\\
   J. Dixon, {``Cohomology and renormalization of  gauge
     theories'', I, II, III}, {\em unpublished preprints, 1976-1977}; \\
  J. Dixon, \cmp{139}{91}{495}\\
  M. Dubois-Violette, M. Henneaux, M. Talon and
  C.-M. Viallet, \pl{B289}{92}{361}\\
  F. Brandt, N. Dragon and M. Kreuzer, \pl{B231}{89}{263}
                                    \np{B332}{90}{224, 250}\\
                 F. Brandt, PhD Thesis (in German),
              {\em University of Hannover (1991), unpublished}.
\end{thebibliography}
\end{document}